\documentclass[12pt]{article}
\usepackage{amsmath}
\usepackage{graphicx,psfrag,epsf}
\usepackage{enumerate}
\usepackage{natbib}
\usepackage{amsmath,amsthm,amsfonts,epsfig,amssymb,natbib,eucal,eufrak}
\usepackage{booktabs}
\usepackage{multirow}
\usepackage{siunitx}
\usepackage{qtree}

\usepackage{float}
\restylefloat{table}
\usepackage{color}

\newtheorem{ex}{Example}
\newtheorem{result}{Result}

\newtheorem{cor}{Corollary}

\newcommand{\blind}{0}

\begin{document}

\def\spacingset#1{\renewcommand{\baselinestretch}%
{#1}\small\normalsize} \spacingset{1}


\if0\blind
{
  \title{\bf Sterrett Procedure for the Generalized Group Testing Problem}
  \author{Yaakov Malinovsky
    \\
    Department of Mathematics and Statistics\\ University of Maryland, Baltimore County, Baltimore, MD 21250, USA\\
}
  \maketitle
} \fi

\if1\blind
{
  \bigskip
  \bigskip
  \bigskip
  \begin{center}
    {\LARGE\bf Title}
\end{center}
  \medskip
} \fi

\bigskip
\begin{abstract}
Group testing is a useful method that has broad applications in
medicine, engineering, and even in airport security control. Consider a finite population of $N$ items, where item $i$ has a probability $p_i$ to be defective.
The goal is to identify all items by means of group testing. This is the generalized group testing problem. The optimum procedure, with respect to the expected total number of tests, is unknown even in case when all $p_i$ are equal. \cite{H1975} proved that an ordered partition (with respect to $p_i$) is the optimal for the Dorfman procedure (procedure $D$), and obtained an optimum solution (i.e., found an optimal partition) by dynamic programming.
In this paper, we investigate the Sterrett procedure (procedure $S$).
We provide close form expression for the expected total number of tests, which allows us to find the optimum arrangement of the items in the particular group.
We also show that an ordered partition is not optimal for the procedure $S$ or even for a slightly modified Dorfman procedure (procedure $D^{\prime}$).
This discovery implies that
finding an optimal procedure $S$ appears to be a hard computational problem. However, by using an optimal ordered partition for all procedures, we show that procedure $D^{\prime}$
is uniformly better than procedure $D$, and based on numerical comparisons, procedure $S$ is uniformly
and significantly better than procedures $D$ and $D^{\prime}$. 
\end{abstract}

\noindent%
 \vfill

\newpage
\spacingset{1.45} 
\section{Introduction}
\label{se:I}
In $1943,$ Robert Dorfman introduced the concept of group testing as a need to administer syphilis tests to millions of individuals drafted into the U.S. army during  World War II.
The test for syphilis was a blood test called the Wassermann test \citep{W1906}. The nice description of the \cite{Dorfman1943} procedure is given by \cite{F1950}:
 {\it "A large number, $N$, of people are subject to a blood test. This can be administered in two ways. (i) Each person is tested separately. In this case $N$ tests are required. (ii) The blood samples of $k$ people can be pooled and analyzed together. If the test is negative, this one test suffices for the $k$ people. If the test is positive, each of the $k$ persons must be tested separately, and all $k+1$ tests are required for the $k$ people. Assume the probability $p$ that the test is positive is the same for all  and that people are stochastically independent.}"\\
Procedure $(ii)$ is commonly referred to as the Dorfman two-stage group testing procedure.

Since the Dorfman work, the group testing has wide spread applications.
The partial list of applications includes blood screening to detect human immunodeficiency virus (HIV),
the detection of hepatitis B virus (HBV), and other diseases
(\cite{G1994}; \cite{BL2010}; \cite{BT2010}; \cite{S2011}), for screening chemical compounds as part of the drug discovery (\cite{Z2001}), DNA screening (\cite{M1999}; \cite{Dh2006}; \cite{CS2016}), quality control in product testing (\cite{SG1959}; \cite{BP1990}), and communication networks (\cite{W1985}).

There is a logical inconsistency in the Dorfman two-stage group testing procedure (procedure {\it D} ).
It is clear that any ``reasonable" group testing plan should satisfy the following property:
``A test is not performed if its outcome can be inferred from previous test results" (\cite{U1960}, p. 50).
The procedure $D$ does not satisfy this property,
since if the group is positive and all but the last person are negative, the last person is still tested.
The modified Dorfman procedure \citep{SG1959} (defined as $D^\prime$) would not test the last individual in this case.
Even though this improvement is small with respect to
the expected number of tests, it has a large impact on optimal design \citep{MA2016}.

\cite{S1957} suggested an improvement of the procedure $D^{'}$ in the following way.
If in the first stage of the procedure $D^{'}$ the group is infected, then in the second stage
individuals are tested one-by-one until the first infected individual is identified.
Then the first stage of procedure is applied to the remaining (non-identified) individuals.
The procedure is repeated until all individuals are identified. The intuition behind Sterrett procedure (procedure $S$) is the following: if $p$ is small, then the probability that two or more subgroups are positive is very small and it will bring the advantage of procedure $S$ over $D^{'}$.
On the other hand, the procedures $D, D^{'}$ are very simple and the second stage can be performed (if there is a need) simultaneously, but procedure $S$ is sequential one-by-one and, therefore, more time consuming. Therefore, practical application depends on a real need.
The recent results, comparisons, and review of these procedures can be found in \cite{MA2016}.

There are only a few fundamental results in binomial group testing that provide insights on the structure of $E(N,p)$, the minimum expected number of tests, which generally is not known. The most important one is due to \cite{U1960}.
Ungar characterized the optimality of any group testing algorithm and proved that if $p\geq p_{U}=(3-5^{1/2})/2\approx 0.38$, then there does not exist an algorithm that is better than individual one-by-one testing, i.e., $\displaystyle E(N,p)=N,\,\,\,\,\text{for}\,\,\,\,\,p\geq p_{U}$.

To the best of our knowledge, the generalized group testing problem was introduced, in the first time, by \cite{NS1973}.  In the motivation problem, the units arrived in the groups of different sizes: group $i$ has size $i$. Therefore, the probability that the group $i$ is defective, (i.e., include at least one defective unit) is $1-q^{i}$, $q=1-p$. Then group $i$ can be treated as a single unit with $\displaystyle p_i=1-q^{i}$.

This work deals with the generalized group testing problem (GGTP): $N$ stochastically independent units, where unit $i$ has the probability $p_i$ ($0<p_i<1$) to be defective.
All units has to be classified as good or defective by group testing. The group testing procedure $A$ is better (not worse) than procedure $B$, if total expected number of tests under the procedure $A$  is less (less or equals) than under the procedure $B$ for any set $\left\{p_1, p_2,\ldots,p_N\right\}$. It is important
to note that even under equal probabilities case an optimum procedure is unknown (for the discussion see \cite{MA2016}).
We assume that the probabilities $\left\{p_1, p_2,\ldots,p_N\right\}$ are known and we can decide about the order in which the people will be tested.

For the GGTP, we redefined the group testing procedures $D,\,D^{'},$ and $S$ to be partition of the finite population of size $N$ into any number of disjoint
groups/subsets. Then procedure $A$ ($A\in \left\{D, D^{\prime}, S\right\}$) is performed on each of these groups.

Ideally, under procedure $A$ ($A\in \left\{D, D^{\prime}, S\right\}$) we are interested in finding an optimal {\it partition} $\displaystyle \left\{m_1,\ldots m_I\right\}$  with $\displaystyle m_1+\ldots+m_I=N$ for some $I\in \left\{1,\ldots,N\right\}$ such that the total expected number of tests is minimal, i.e.,

\begin{align}
\label{eq:Goal}
&
\displaystyle \left\{m_1,\ldots m_I\right\}=\arg\min_{n_1,\ldots,n_J}
E_{A}\left(n_1,n_2,\ldots,n_J\right),\nonumber\\
&
\text{subject to}\,\,\,\,\,\, \sum_{i=1}^{J}n_i=N,\,\,\, J\in\left\{1,\ldots,N\right\},
\end{align}
where  $\displaystyle E_{A}\left(n_1,n_2,\ldots,n_J\right)=E_{A}\left(1:n_1\right)+E_{A}\left(1:n_2\right)+\ldots+E_{A}\left(1:n_J\right)$, and $\displaystyle E_{A}\left(1:n_j\right)$ is the total expected number of tests (under procedure $A$) in the group of size $n_j$.

It is important to note that from Ungar's fundamental result (discussed above), it follows that if $p_i\geq p_{U}=(3-5^{1/2})/2\approx 0.38$ for any $i=1,\ldots,N$ then
does not exist a group testing algorithm which is better than a one-by-one individual testing. In this case, the solution of \eqref{eq:Goal} is $\displaystyle n_i=1,\, i=1,\ldots,I,\, I=N$.

Potentially, the optimization problem \eqref{eq:Goal} can be solved under brutal search, i.e., evaluating the right-hand side of the first equation in \eqref{eq:Goal} for any possible partition and then choosing an optimal one.
But this task is a hard computational problem and it is impossible to perform because the total number of possible partitions of a set of size $N$ is the Bell number
$\displaystyle B(N)=\left\lceil \frac{1}{e}\sum_{j=1}^{2N}\frac{j^N}{j\,!}\right\rceil$
(\cite{B1934},\,\cite{BB2006}), which grows exponential with $N$. For example, $B(3)=5, B(5)=52, B(10)=115,975, B(13)=27,644,437$.

However, for some cost functions (in our case the cost function is $E_{A}\left(n_1,n_2,\ldots,n_J\right)$) a solution can be obtained with polynomial in $N$ computation cost (see, e.g., \cite{H1975}; \cite{T1979}; \cite{H1981}; \cite{Ch1982}).

In the group-testing setting, \cite{H1975, H1981} proved that under procedure $D$ an optimal partition is an ordered partition  (i.e., each pair of subsets has the property that the numbers in one subset are all great or equal to every number in the other subset). The total number of ordered partition of a set of size $N$ is $2^{N-1}$. The brutal search among all $2^{N-1}$ possibilities is still a computational intractable problem \citep{GJ1979}.
But in this case, \cite{H1975} showed the existence of a polynomial time algorithm with the computational effort $O(N^2)$.
This algorithm is the dynamic programming algorithm \citep{B1957}.

\cite{BT2010} applied procedure $S$ to chlamydia and gonorrhea testing in Nebraska. For each covariate (gender and specimen) combination, they randomly
assigned the individuals to groups of constant size. Then they estimated the individual risk probabilities $p_i$ in each group using a logistic model.
The procedure $S$ was applied to each group. The individuals' insight of the particular group were tested (if necessary) according to the decreasing order of $p_i$,
i.e., first, the individual with the largest $p_i$ value, next the second largest, and so on. They assumed that the diagnostic tests are not error-free.
Generally, this assumption is realistic and has to be taken into account. We do not attempt to investigate erroneous tests in the current work and discuss this as an important direction for the future investigation.

To the best of our knowledge, no optimality result is available to the other group testing procedures for the generalized problem, including procedure $S$.
It can be explained by the fact that the close form expression for the
$\displaystyle E_{S}\left(1:k\right)$ was not available and it is essential.

In this work, we find the closed-form expression and also show how items should be arranged in the particular group such that $\displaystyle E_{S}\left(1:k\right)$ will be minimal. We also show that for the both procedures $D^{'}$ and $S$ an optimal partition is not ordered.
Therefore, there is no available optimum solution here with polynomial in $N$ computation effort, and it seems to be a computationally hard problem.
But, despite this negative answer, the performance
of the procedure $S$ under an optimal ordered partition (which is not optimal, but can be obtained with $O(N^2)$ computational effort) is significantly better than the performance of an optimal procedure $D$.
The comparisons with procedure $D^{'}$ will be made also.

\section{Basic Characteristics of the Procedures}
\subsection{Procedure $S$ }
In order to investigate the Sterrett procedure, we have to find the expected number of tests, in the group of size $k$, under this procedure.
First, we present below two examples for the particular cases $k=2$ and $k=3$. It will help to develop the intuition for the general case.
\\
\begin{ex}[Group of size $k=2$]
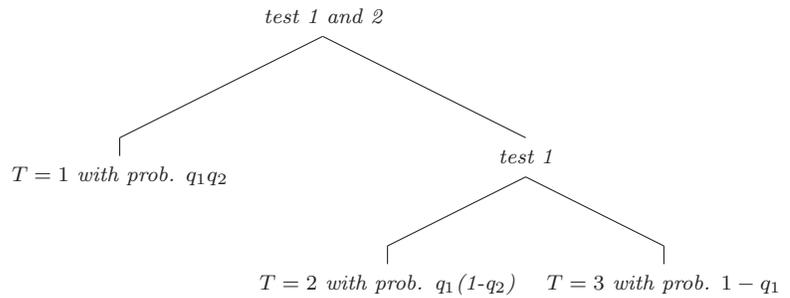
\scriptsize
\Tree [.{test 1 and 2} [{$T=1$ with prob. $q_1 q_2$} ] [.{test 1} [{$T=2$ with prob. q_1(1-q_2)} ] [{$T=3$ with prob. $1-q_1$}  ] ] ]
\end{ex}
The tree above represents Sterrett procedure for the group of size $k=2$.
Starting from the root of the tree with two individuals, we perform the procedure and stop at terminal node when all (here 2) individuals are identified.
The left branch of the tree represents the negative test result, and the right branch represents the positive test result.
The total number of tests $T$ with the corresponding probabilities are presented in the terminal nodes.
From the tree, we obtain the expected total number of tests.
\begin{align}
\label{eq:2}
&
E_{S}\left(1:2\right)=1q_1q_2+2q_1(1-q_2)+3(1-q_1)=3-q_1-q_1q_2.
\end{align}
From Equation \eqref{eq:2}, it follows that arranging the probabilities in decreasing order $\displaystyle q_1\geq q_2$  we obtain the minimum value of
$\displaystyle E_{S}\left(1:2\right)$.

The next example for the case $k=3$ will help us to approach the general case.

\begin{ex}[Group of size $k=3$]
\noindent
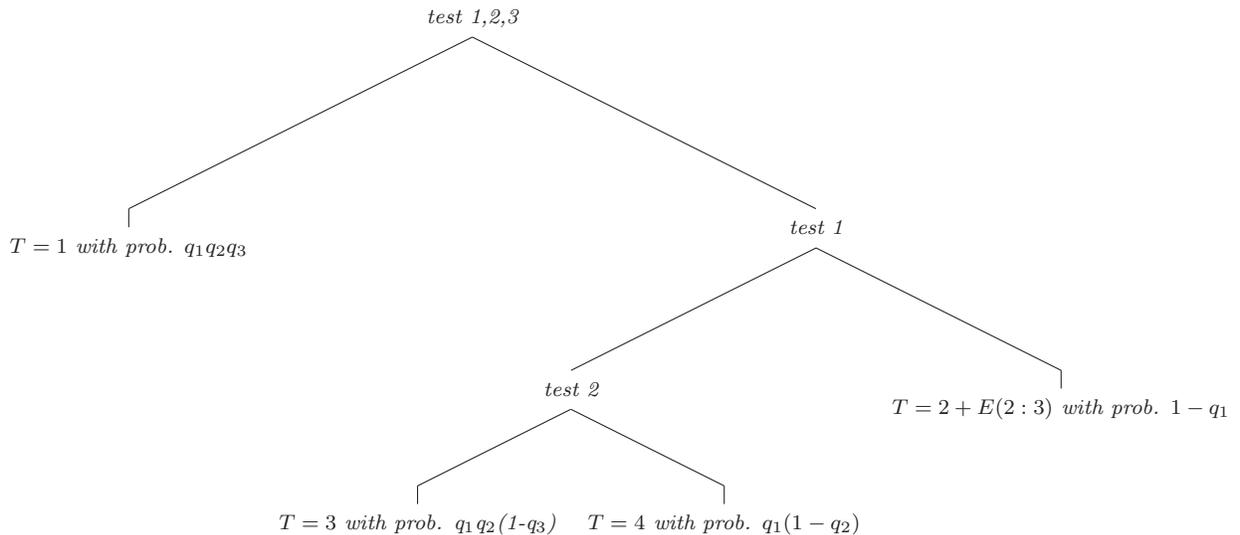
\begin{center}
\scriptsize
\Tree [.{test 1,2,3} [{$T=1$ with prob. $q_1q_2q_3$} ] [.{test 1} [.{test 2} [{ $T=3$ with prob. q_1q_2(1-q_3)}  ][{$T=4$ with prob. $q_1(1-q_2)$}  ] ] [{$T=2+E(2:3)$ with prob. $1-q_1$} ]  ] ]
\end{center}
\end{ex}
The expected total number of tests is
\begin{align}
\label{eq:3}
&
E_{S}\left(1:3\right)=1q_1q_2q_3+3q_1q_2(1-q_3)+4q_1(1-q_2)+\left(2+E(2:3)\right)(1-q_1)\nonumber \\
&
=5-q_1-q_2-q_2q_3-q_1q_2q_3.
\end{align}
From Equation \eqref{eq:3}, it follows that arranging the probabilities in the order  $q_2\geq q_1 \geq q_3$  we obtain the minimum value of
$\displaystyle E_{S}\left(1:3\right)$.

These two examples are induction steps for the general case which is presented below.
\begin{result}
\label{res:main}
\begin{itemize}
\item[(i)] The total expected number of test under generalized Sterrett procedure in the group of size $k$ $(k\geq 1)$ is
\begin{align}
\label{ex:main}
&
E_{S}\left(1:k\right)=(2k-1)-\left[(q_1+q_2+\ldots+q_{k-1})+q_{k-1}q_{k}+q_{k-2}q_{k-1}q_{k}+\ldots+q_1q_2\ldots q_k\right].
\end{align}
\item[(ii)]
For $k\geq 2$, the minimum of $\displaystyle E_{S}\left(1:k\right)$ is obtained under the order
\begin{align}
\label{res:order}
\displaystyle q_{k-1}\geq q_{k-2}\geq \ldots \geq q_1 \geq q_{k}.
\end{align}
\end{itemize}
\end{result}
For the proof of Result \ref{res:main}, see Appendix \ref{A:aa}.

\begin{cor}
If $\displaystyle p_1=\ldots=p_N\equiv p$,  then
\begin{align}
\label{eq:HS}
&
E_{S}\left(1:k\right)=2k-(k-2)q-\frac{1-q^{k+1}}{1-q},\,\,\,\,\,\,\,\,\,\,\,\,\, k\geq 1.
\end{align}
\end{cor}
It is an interesting to note that in the original work of \cite{S1957} the closed form \eqref{eq:HS} was not provided.
For the discussion, please see \cite{MA2016}.

\subsection{Procedures $D$ and $D^{'}$}
In this section, we will compare the procedure $S$ with the procedures $D$ and $D^{'}$.
For this purpose, we need to present some characteristics for the procedures $D$ and $D^{'}$.
Below we present the expected total number of tests in the group of size $k$ for the procedures $D$ and $D^{'}$.
\\
\\
\noindent
{\it Procedure D}\\
For $k\geq 2,$ the total number of tests is $1$ with probability $\displaystyle \prod_{i=1}^{k}q_i$ and $k+1$ with probability  $\displaystyle 1-\prod_{i=1}^{k}q_i$.
Therefore,
\begin{equation}
\label{eq:D}
E_{D}(1:k)=1+\left(k-k\prod_{i=1}^{k}q_i\right)1_{\{k\geq 2\}}.
\end{equation}
\\
\\
\noindent
{\it Procedure $D^{'}$}\\
For $k\geq 2,$ the total number of tests is $1$ with probability $\displaystyle \prod_{i=1}^{k}q_i$, $k$ with probability  $\displaystyle \prod_{i=1}^{k-1}q_i(1-q_k)$,
and $k+1$ with probability $\displaystyle 1-\prod_{i=1}^{k}q_i-\prod_{i=1}^{k-1}q_i(1-q_k)$. Therefore,
\begin{equation}
\label{eq:D'}
E_{D^{'}}(1:k)=1+\left(k-k\prod_{i=1}^{k}q_i-\prod_{i=1}^{k-1}q_i(1-q_k)\right)1_{\{k\geq 2\}}.
\end{equation}

From Equations \eqref{eq:D} and \eqref{eq:D'}, we obtain three important observations:
\begin{itemize}
\item[{(i)}]
In the procedure $D$, the order in which items are tested in the particular group is not important, the value of $\displaystyle E_{D}(1:k)$ is the same for any permutation.
It is not so under procedure $D^{'}$. In order to minimize $\displaystyle E_{D^{'}}(1:k)$, the last tested item, say item $k$, should correspond to the smallest $q_i$
in the group.
\item[{(ii)}]
From Equations \eqref{eq:D} and \eqref{eq:D'}, it obviously follows that $$\displaystyle E_{D^{'}}(1:k)\leq E_{D}(1:k).$$
\item[{(iii)}]
From the definitions of the procedures $\displaystyle D^{'}$ and $S$, it follows that they are equivalent for $k=2$, i.e.,  $\displaystyle E_{D^{'}}(1:2)=E_{S}(1:2)$.
It  also follows from Equations \eqref{eq:2}  and \eqref{eq:D'}.
\end{itemize}

\section{Finding an Optimal Procedures $D^{\prime}$ and $S$  Appears to be a Hard Computational Problem }
It was mentioned in the Introduction that for the arbitrary cost function, the solution of \eqref{eq:Goal} is computationally hard problem (see, e.g., \cite{GJ1979}).

We also mentioned that \cite{H1975} had proved that an optimal ordered partition is an optimal for the procedure $D$.
In the conclusions to his work, \cite{H1975} wrote: {``\it A Dorfman procedure has the property that a unit is classified as a defective only after an individual test finds it so. There are procedures which allow a unit to be classified as defective by deduction, e.g., if a group of size $g$ is found defective and subsequent tests find g - 1 members of this group good, then the remaining member is classified as a defective without further testing. The savings of cost due to this possible deduction is small, but the mathematics to obtain optimal procedures would be extremely complicated. Therefore, we do not pursue this line.}"
It is clear that procedures $D^{\prime}$ and $S$ satisfies the ``{\it deduction}" property mentioned by Hwang.
Even though this improvement is small with respect to the expected number of tests, it has a large impact on optimal design. It was explained in \cite{MA2016}
for the case $\displaystyle p_1=\ldots=p_N$.

The following example shows that the optimal ordered partition is not optimal for the  procedures $D^{\prime}$ and $S$.

\begin{ex}
Take $N=4$ and $\left\{q_1, q_2, q_3, q_4\right\}=\left\{0.6, 0.6, 0.99,0.99\right\}.$
There are $8$ possible ordered partitions. The optimum one among them can be found using \eqref{ex:main} and \eqref{eq:D'} through direct calculation or using dynamic programming algorithm (Appendix \ref{B:bb}  ).
Such optimum ordered partition is the same for both $S$ and $D^{\prime}$:\, $\displaystyle \left\{0.6\right\} \cup  \left\{0.6, 0.99, 0.99\right\}$
with the corresponding $E_{S}=2.83794$ and $E_{D^{\prime}}=2.8438$.\\
Now, consider the following unordered partition $\displaystyle \left\{0.6, 0.99\right\}\cup  \left\{0.6, 0.99\right\}$.
For this partition, we have $\displaystyle E_{S}=E_{D^{'}}=2.832$ showing non-optimality of the optimal ordered partition for both procedures.
\end{ex}

The above example shows that finding an optimal solution for \eqref{eq:Goal} seems to be computationally hard problem under procedures $D^{\prime}$ and $S$.

Another interesting observation is the following. As we already mentioned $\displaystyle E_{D^{'}}(1:2)=E_{S}(1:2)=3-q_1-q_1q_2.$
For $N=4$ without loss of generality, let us assume $\displaystyle q_1\geq q_2 \geq q_3 \geq q_4$. Suppose that we perform procedure $S$ (or $D^{'}$) to the following ordered partition $\displaystyle \left\{q_1,\,q_2\right\}\cup \left\{q_3,\,q_4\right\}.$  By interchanging $q_2$ with $q_3$ (and creating unordered partition), we always reduce the expected total number of tests.

\subsection{Comparisons Among the Procedures}
\subsubsection{Comparisons}
The following result is useful for the comparison.

\begin{result}
\label{res:Dp}
Denote $\displaystyle \left\{m_{1},\ldots,m_{I}\right\}$ an optimal partition under procedure $D$, and $\left\{m^{'}_{1},\ldots,m^{'}_{I^{\prime}}\right\}$
an optimal ordered partition under procedure $D^{\prime}$. From Equations \eqref{eq:D}, \eqref{eq:D'}, and \cite{H1975,H1981} (which proved that
an optimal partition under procedure $D$ is ordered), the following relations immediately follows:
\begin{equation}
E_{D^{\prime}}\left(m^{'}_{1},\ldots,m^{'}_{I^{\prime}}\right)\leq E_{D^{\prime}}\left(m_{1},\ldots,m_{I}\right)\leq E_{D}\left(m_{1},\ldots,m_{I}\right).
\end{equation}
\end{result}

Denote $\left\{m^{*}_{1},\ldots,m^{*}_{I^{*}}\right\}$ an optimal ordered partition under Sterrett procedure. We are going to compare
$\displaystyle E_{S}\left(m^{*}_{1},\ldots,m^{*}_{I^{*}}\right)$ with $\displaystyle E_{D^{\prime}}\left(m^{'}_{1},\ldots,m^{'}_{I^{\prime}}\right)$.

The comparisons will be done in the following manner.
We generate the vector $\displaystyle p_1,p_2,\ldots,p_{100}$ from Beta distribution with parameters $\displaystyle \alpha=1, \beta>0$ such that
$\displaystyle \frac{1-p}{p}=\beta$, i.e., expectation equals to $p$, and standard deviation equals to $\displaystyle p \sqrt{\frac{1-p}{1+p}}$ (second column of Table \ref{t:1}).
We repeat this process $M=1000$ times for each value of $p$.
Each time a solution of the optimization problem \eqref{eq:Goal} under ordered partition restriction for procedures $D$, $D^{\prime}$, and $S$ was obtained using dynamic programming (DP) algorithm, which is presented in Appendix \ref{B:bb}. From DP algorithm we also obtained
the expected total number of tests for an optimal ordered partition in procedures $D$, $D^{\prime}$, and $S$.

In the following table,  we present a mean (over 1000 repetitions) of the expected total number of tests under these three procedures  and in the bracket we present the standard deviation of the mean.

\begin{table}[H]
\caption{Comparison of the procedures $D$, $D^{\prime}$, and $S$ under optimal ordered partitions}
\label{t:1}
\small
\begin{center}
  \begin{tabular}{llSSSSS}
    \toprule
    \multirow{2}{*}{p} &
    \multirow{2}{*}{std} &
      \multicolumn{4}{c}{$N=100$ } \\
      &{}& {$D$} & {$D^{'}$} & {$S$} & {$H(P)$}\\
      \midrule
    0.001& 0.001   &{5.7519 (0.009)}&{5.7383(0.009)}&{3.7453(0.006)}&{1.0807(0.003)}           &        \\
    0.01&  0.0099&{17.47(0.0279)}  &{17.345(0.0271)}&{13.121(0.0235)}&{7.4735(0.0191)}&\\
    0.05&0.0476& {38.102(0.0581)}&{37.095(0.0567)}&{31.801(0.0565)}&{25.653(0.0584)}&\\
    0.10&0.0905&{52.866(0.0762)}&{50.758(0.0695)}&{46.105(0.0736)}&{40.855(0.0780)}&\\
    0.20&0.1633& {70.266(0.0826)}&{67.536(0.0762)}&{64.33(0.0828)}&{60.11(0.0878)}&\\
    0.30&0.2201& {80.067(0.0802)}  &{77.598(0.0772)}&{75.358(0.0844)}&{70.303(0.0874)}&\\
   \bottomrule
  \end{tabular}
  \end{center}
  \end{table}

As we see in this table, there is uniform dominance of the procedure $S$ over $D$ and $D^{\prime}$.
For the small values of $p,$ the saving can be very significant. For example, if the mean value equals to $0.01$
then the procedure $S$ requires $13.121$ tests per 100 individuals, versus $17.47$ under procedure $D^{\prime}$.

\subsubsection{Lower Bound}
The connection between group testing and information theory was described by \cite{SG1959} and the connection with codding theory by \cite{S1960}
for the case $p_i=p$ for all $i$. For the comprehensive discussion, see \cite{K1973}.

These results can be extended to the generalized group testing problem. For example, in the case $N=2$ with $q_1>\frac{1}{2}$ and $q_2>\frac{1-q_1}{q_1}$
both procedures $D^{\prime}$ and $S$ are preferred to individual testing and optimal. The optimality follows from the fact that both procedures are equivalent to the optimal prefix Huffman code \citep{H1952}
with the expected length $L(N),\, N=2$.
For the $N\geq 3,$ the optimum group testing strategy does not attain $L(N)$. It was mentioned by  \cite{S1967} for the case $p_i=p$ for all $i$
and, therefore, it holds for the generalized group testing problem.
But for the GGTP, an optimal prefix Huffman code with the expected length $L(N)$ serves as a theoretical lower bound
for the unknown optimal GGTP \citep{NS1973}.

\noindent
The explicit form of $L(N)$ is available only for the special cases with particular restrictions on the probabilities $p_1,\ldots,p_N$ \citep{KL1972, K1973, B1974}.
It is known that in general, the complexity of calculation of $L(N)$ is $O\left(2^{N}\log_{2}(2^N)\right)$ due to the sorting effort.
Therefore, even for small $N$, obtaining the exact value of $L(N)$ is impossible.
A well-known result in information theory (Noiseless Coding Theorem; see e.g., \cite{K1973}, \cite{CT2006}) provides the information theory bounds for $L(N)$:
\begin{equation}
H(P)\leq L(N) \leq H(P)+1,
\end{equation}
where $\displaystyle P=\left(p_1,\ldots,p_N\right)$ and $\displaystyle H(P)=\sum_{i=1}^{N}\left\{p_i\log_{2}\frac{1}{p_i}+q_i\log_{2}\frac{1}{q_i}\right\}$
is the Shannon entropy. This value $H(P)$ can be used as the information lower bound for an optimal group testing procedure.

For the numerical example in Table \ref{t:1}, the entropy $H(P)$ was calculated for each simulation (total 1000) and mean value (standard deviation of the mean in the brackets) was reported in the last column.

\section{Conclusions and Discussion}
For all procedures discussed here, finding an optimal solution is equivalent to finding an optimal partition.
In general, an optimum partition problem is a hard computational problem. This is why the optimal solution for the generalized group testing problem is only
available for the Dorfman procedure \citep{H1975, H1981}. In this work we try to shed light on the additional procedures.
It was shown that finding an optimal partition under procedures $S$ and $D^{\prime}$ appears to be a hard computational problem.
However, using an optimal ordered partition for these procedures, with the computational cost $O(N^2)$, will substantially improve the optimal procedure $D$.
As a byproduct, the closed-form expression for the expected total number of tests  was obtained for the Sterrett
procedure. The investigation of the general class of nested procedures, to which both $D^{\prime}$ and $S$ belong, is a natural next step for future work.
Another direction for future investigation is to relax the assumption of error-free tests. Generally, in this case, the expected total number of tests
cannot be used as the only criterion for comparison among group testing procedures and additional criteria have to be considered.
\cite{MAR2016} suggested a possible criterion where procedure $D$ was investigated for the equal probabilities case.
Much more work has to be done for generalized group testing problems.

\section*{Acknowledgement}
I thank James B. Orlin for the detailed explanations related to his article, and I also
thank Shelemyahu Zacks for the discussions and comments on the article.
The author thanks the editor, associate editor, and two referees for their time, advice, and constructive comments.
The author also thanks Mattson Publishing Services for editing the article.

\bigskip




\section*{Appendix} \appendix

\section{Proof of Result \ref{res:main}}
\label{A:aa}
\begin{proof} The proof of both parts (i) and (ii) is by induction on $k$.
\begin{itemize}
\item[(i)]
For $k=2,$ the result holds (Example 1). Assume that \eqref{ex:main} is correct for $k-1$. We prove that it is correct for $k$.  Let $1_j$ be indicator function, equals 1 if the first positive individual identified is the individual $j$ ($j=1,\ldots,n$). Let $1_0$ be the indicator function that there are no positive individuals in the group. Denote $T_k$ be the total number of the tests in the group of size $k$.
It is clear that $$T_k=T_k1_{0}+T_k1_{1}+\ldots+T_k1_{k}.$$
We have $\displaystyle E\left(T_k 1_0\right)=q_1\cdots q_k,$\,\,\,\,
$\displaystyle E\left(T_k 1_{k}\right)=q_1\cdots q_{k-1}(1-q_{k})k,$\,\,\,\, $\displaystyle E\left(T_k 1_{k-1}\right)=q_1\cdots q_{k-2}(1-q_{k-1})(k+1)$, and\,\,\,\,
$\displaystyle E\left(T_k 1_{j}\right)=q_1\cdots q_{j-1}(1-q_{j})\left(1+j+E(j+1:k)\right),\,\,j=1,\ldots,k-2,$ where $q_{1}\cdots q_{0}\equiv 1$.\\
Combining all together, we have
\begin{align}
\label{eq:main}
&
E\left(1:k\right)=E\left(T_k\right)=\sum_{j=0}^{k}E\left(T_k 1_{j}\right)=q_1\cdots q_k+q_1\cdots q_{k-1}(1-q_{k})k+q_1\cdots q_{k-2}(1-q_{k-1})(k+1)\nonumber\\
&
+
\sum_{j=1}^{k-2}q_1\ldots q_{j-1}(1-q_j)\left(1+j+E\left(j+1:k\right)\right)=\nonumber\\
&
-q_1\cdots q_k\left(k-1\right)-q_1\cdots q_{k-1}+q_1\cdots q_{k-2}\left(k+1\right)
+
\sum_{j=1}^{k-2}q_1\ldots q_{j-1}(1-q_j)\nonumber\\
&
\left(1+j+\left\{2(k-j)-1-\left[q_{j+1}+\ldots+q_{k-1}\right]-\left[q_{j+1}\cdots q_k+\ldots+q_{k-1}q_{k}\right]\right\}\right)\nonumber\\
&
=(2k-1)-q_1\cdots q_k\left(k-1\right)-q_1\cdots q_{k-1}-q_1\cdots q_{k-2}-q_1\cdots q_{k-3}-\ldots-q_1q_2-q_1\nonumber
\\
&
-
\sum_{j=1}^{k-2}q_1\ldots q_{j-1}(1-q_j)
\left\{\left[q_{j+1}+\ldots+q_{k-1}\right]+\left[q_{j+1}\cdots q_k+\ldots+q_{k-1}q_{k}\right]\right\}\nonumber\\
&
=(2k-1)-q_1\cdots q_k\left(k-1\right)-q_1\cdots q_{k-1}-q_1\cdots q_{k-2}-q_1\cdots q_{k-3}-\ldots-q_1q_2-q_1\nonumber\\
&
+q_1\cdots q_k\left(k-2\right)+q_1\cdots q_{k-1}+q_1\cdots q_{k-2}+q_1\cdots q_{k-3}+\ldots+q_1q_2\nonumber\\
&
-
\left(q_1+\ldots+q_{k-1}\right)-\left(q_{k-1}q_{k}+q_{k-2}q_{k-1}q_{k}+\ldots+q_{2}\cdots q_{k-1}q_{k}\right)\nonumber\\
&
=(2k-1)\nonumber
\\
&
-\left[(q_1+q_2+\ldots+q_{k-1})+q_{k-1}q_{k}+q_{k-2}q_{k-1}q_{k}+\ldots+q_2\cdots q_k+q_1q_2\cdots q_k\right].
\end{align}

\item[(ii)]
Recall that the people are tested in the order $1,2,\ldots,k-1,k$ with corresponding probabilities  $q_1,q_2,\ldots,q_{k-1},q_{k}$.
The set of these probabilities are available before we start the test process. Therefore, we are free to decide the order of the probabilities
from this set, which we will assign to the persons $1,2,\ldots,k$
in order to minimize the left-hand side of the equation \eqref{eq:main}. The the optimal assignments to minimize the left-hand side of the equation \eqref{eq:main} is equivalent to finding an assignment that maximizes
\begin{equation}
\label{eq:k}
e_{1:k}=(\underbrace{q_1+q_2+\ldots+q_{k-1}}_{S_{1:{k-1}}})+\underbrace{q_{k-1}q_{k}+q_{k-2}q_{k-1}q_{k}+\ldots+q_2\cdots q_k}_{R}+q_1q_2\cdots q_k.
\end{equation}

For $i<k$, denote $G_{i:k}=\left\{q_i,\ldots,q_k\right\}.$
We start with $k=2$. We have to decide how to arrange the values of $G_{1:2}=\left\{q_1, q_2\right\}$ in order to maximize
$q_1+q_1q_2$. It is obvious that we have to choose $q_1\geq q_2$. Assume that \eqref{res:order} holds for the group of size $k-1$ with $\displaystyle G_{2:k}=\left\{q_2,\ldots,q_k\right\}$, i.e.,
\begin{equation}
\label{eq:ind}
q_{k-1}\geq \ldots \geq q_3\geq q_2 \geq q_k,
\end{equation}
with the corresponding
\begin{align}
\label{eq:cor}
&
e_{2:k}=(\underbrace{q_2+\ldots+q_{k-1}}_{S_{2:{k-1}}})+q_{k-1}q_{k}+q_{k-2}q_{k-1}q_{k}+\ldots+q_2\ldots q_k.
\end{align}

We prove that it is true for the group of size $k$. Comparing \eqref{eq:cor} with \eqref{eq:k}
we see that $S_{1:{k-1}}=q_1+S_{2:{k-1}}$, $R$ does not include $q_1$, and the last term $q_1q_2\cdots q_k$ in the equation \eqref{eq:k}
is the constant for the given set of the probabilities. Therefore, to maximize \eqref{eq:k} we must choose $q_1=\min_{1\leq j\leq k-1}\left\{q_j\right\}$. Combining this with the induction step \eqref{eq:ind} we conclude that we have to decide between two alternatives
$q_{k-1}\geq \ldots \geq q_3\geq q_2\geq q_1 \geq q_k$  and $q_{k-1}\geq \ldots \geq q_3\geq q_2\geq q_k\geq q_1$. Direct calculation shows that the first one will maximize \eqref{eq:k} (or, as an alternative minimize \eqref{ex:main}).

\end{itemize}
\end{proof}

\section{Dynamic Programming Algorithm for Finding an Optimal Ordered Partition}
\label{B:bb}
We need some additional natation to present this well-known algorithm (see, e.g., \cite{H1975}; \cite{Ch1982}).
Let $\displaystyle U_{N}=\left\{u_1,u_2,\ldots,u_N\right\}$ be the population subject to classification. Each unit $u_i$ has the probability $p_i$ to be defective and the probability $q_i=1-p_i$ to be good. Without loss of generality, assume that the units $u_1,\ldots,u_N$ are labeled such that
$\displaystyle 0<p_1\leq p_2\leq \ldots \leq p_N <1.$  The goal is to find an optimal ordered partition under procedure $A$, which is a solution of optimization
problem  \eqref{eq:Goal} under ordered partition restriction.
Denote the cost function $C_{A}\left(U_N\right)$ as a expected total number of tests for the population $U_N$ under an optimum ordered partition using procedure $A$.
DP algorithm solution is
\begin{equation}
\label{eq:11}
C_{A}(U_0)=0,\,\,\, C_{A}(U_1)=1,\,\,\,
C_{A}\left(U_{k}\right)=\min_{0\leq i \leq k-1}\left\{E_{A}\left(U_{k}-U_{i}\right)+C_{A}\left(U_i\right)\right\},\,\,\,\,2\leq k \leq N,
\end{equation}
where $\displaystyle E_{A}\left(U_{k}\right)$ is the expected total number of tests of testing group $U_{k}$ under procedure $A$,
$\displaystyle U_0=\emptyset$.

{}

\end{document}